\documentclass[aps,prl,twocolumn,groupedaddress]{revtex4}

\usepackage[dvips]{graphicx}

\begin{document}

\title{Charm Coalescence at relativistic energies}

\author{
A.P. Kostyuk$^{1,2}$,
M.I. Gorenstein$^{1,2}$,
H. St\"ocker$^{1}$,
and
W. Greiner$^{1}$
}

\affiliation{$^1$ Institut f\"ur Theoretische Physik, 
J.W. Goethe-Universit\"at, Frankfurt am Main, Germany}

\affiliation{$^2$ Bogolyubov Institute for Theoretical Physics,
Kyiv, Ukraine}

\date{\today}

\begin{abstract}
The $J/\psi$ yield at midrapidity at the top RHIC 
(relativistic heavy ion collider) energy 
is calculated 
within the statistical coalescence model, which assumes 
charmonium formation at the late stage of the reaction
from the charm quarks and antiquarks created earlier
in hard parton collisions. The results are 
compared to the new PHENIX data and to predictions of the 
standard models, which assume formation of  charmonia exclusively 
at the initial stage of the reaction and 
their subsequent suppression.
Two versions of the suppression scenario are considered.
One of them assumes gradual charmonium suppression 
by comovers, while the other one supposes that 
the suppression sets in abruptly due to quark-gluon plasma 
formation. Surprisingly, both versions give very
similar results. In contrast, the statistical coalescence 
model predicts a few times larger $J/\psi$ yield in the most central
collisions.
\end{abstract}


\maketitle

A study of open and hidden charm production in nucleus-nucleus 
($A+A$) collisions at RHIC (relativistic heavy ion collider) BNL 
is expected to shed light upon an  
important physical question of the space-time history of the
charmonium formation.
The standard ``suppression'' approach
is based on the idea of Matsui and Satz \cite{MS}:
charmonia are formed at the early stage of $A+A$ reaction,
the further evolution leads exclusively to
their suppression due to interaction with initial
nucleons from the colliding nuclei, secondary comoving hadrons,
and/or deconfined medium.

The idea of thermal $J/\psi$ production \cite{gago1} triggered the 
development of
an alternative charmonium formation scenario,
the statistical coalescence model (SCM) \cite{BMS,We}.
Hidden charm
mesons are assumed to be created at hadronization near the point of chemical
freeze-out due to coalescence of charm quarks $c$ and antiquarks 
$\bar{c}$ produced at the initial stage. The distribution of $c$'s
and $\bar{c}$'s over different open and hidden charm species is 
given by the laws of equilibrium statistical mechanics.
The SCM describes remarkably well the centrality dependence 
of the $J/\psi$ yield \cite{SPS}  as well as the transverse
spectra \cite{Bug} at SPS. 

A combination of the standard and SCM approaches  \cite{GR} as well
as a nonthermal $c\bar{c}$ coalescence \cite{Thews}
have been also considered.

The preliminary RHIC data \cite{data_prel} on the $J/\psi$
rapidity density and its centrality dependence in
Au+Au collisions at $\sqrt{s}=200$~GeV from 
the PHENIX Collaboration
have been already discussed  in Refs.\cite{theory1,theory2,theory3}.
The final data \cite{data}, which became available 
recently, differ essentially from the preliminary ones.
Because of low statistics, the data have rather large errorbars.
The most likely value of $J/\psi$ yield  in  the most central collisions 
is not reported. Instead, only 90\%  \  confidence level upper limit is given.
Still, the data appeared to be able to exclude the most 
extreme versions of the nonthermal coalescence scenarios \cite{Thews}.

The aim of the present paper is to check whether SCM 
is tolerated by the data and compare its predictions to the standard 
suppression models.

Let two nuclei $A$ and $B$ collide at impact parameter $b$.
The number of produced $J/\psi$ mesons is given within the 
standard scenario by \cite{Kharzeev}
\begin{equation}
\langle J/\psi \rangle_{AB(b)}
=  \sigma^{NN}_{J/\psi}
A B \int d^2 s T_A(|\vec{s}|) T_B(|\vec{s}-\vec{b}|)
S(\vec{b},\vec{s}), \label{Njpsi}
\end{equation}
where $\sigma^{NN}_{J/\psi}$ is the cross section of $J/\psi$
production in nucleon-nucleon ($N+N$) collisions,
 $T_{A(B)}$ is the nuclear thickness function related to the
nucleon density in the nucleus, and $S(\vec{b},\vec{s}) < 1$ is a factor
responsible for the $J/\psi$ suppression.

At the very initial stage, charmonia experience  absorption,
$S=S^{abs}$, by sweeping
nucleons of the colliding nuclei (see, for instance, 
Refs. \cite{Kharzeev,Capella}). Bound $c\bar{c}$ states are
assumed to be absorbed in the so-called preresonance
state, before the final hidden charm mesons are formed.
The absorption cross section is therefore 
taken to be the same for all charmonia. The value 
$\sigma_{abs}=4.4$~mb \cite{Cortese} follows from
the most recent SPS data analysis and is close to 
the theoretical prediction of Ref. \cite{Gerland}.
We assume that the same value of $\sigma_{abs}$ prevails also
at RHIC energies.

Those charmonia that survive normal nuclear suppression are
subjected to the comover \cite{Capella,comover} or quark-gluon
plasma (QGP) suppression \cite{Blaizot}.
Both suppression scenarios describe
successfully the centrality dependence of the $J/\psi$ yield in
Pb+Pb collisions at the SPS.

In the comover approach, an additional suppression factor
appears: $S=S^{abs} S^{co}$ \cite{Capella}.
The suppression factor $S^{co}$ depends on the density of comovers
and on an {\it effective} cross section
$\sigma_{co}$ of $J/\psi$
dissociation by comovers (averaged over all comover species and
all charmonium states contributing to the $J/\psi$ yield through
their decays and also over particle momenta in the medium).
The value $\sigma_{co}=0.65$~mb
\cite{Capella_new} from fits of new SPS NA50 data  \cite{Ramelo}
will be used in our analysis.
%Extrapolating the comover model
%from the SPS  to RHIC energies, 
We assume that 
the value of $\sigma_{co}$ remain the same also at RHIC.
The charmonium suppression at RHIC becomes, however, stronger,
due to the higher comover density.

%To calculate the suppression $S^{co}(\vec{b},\vec{s})$ by  comover
%hadrons,  one needs a value of the comover density in the
%transverse plane per unit rapidity,
%$n_{H}(\vec{b},\vec{s})$,
%and the effective cross section of charmonia dissociation
%due to interaction with comover hadrons. The averaged over all
%charmonia+hadron reactions value of $\sigma_{H}=0.65$~mb
%\cite{Capella_new} based on fitting the SPS NA50 data  \cite{Ramelo}
%will be used in our analysis.

There are two reasons for increasing the comover density at RHIC relative
to SPS.  
The multiplicity
of produced secondary hadrons per unit rapidity interval at midrapidity
increases by a factor of about 1.5
from $\sqrt{s}=17$ GeV to $\sqrt{s}=200$ GeV
already in elementary nucleon-nucleon collisions.
Additionally, the deviations from the wounded nucleon
model becomes stronger at higher energies. This
increases the comover
density in central nucleus-nucleus collisions.
 The centrality dependence of the number
of light-flavored hadrons per unit pseudorapidity interval
in Au+Au collisions at RHIC can be parametrized as \cite{KhN}
\begin{equation}\label{nch_AA}
\left. \frac{d N_{h}^{\mbox{\footnotesize AuAu}}}{d y} \right|_{y=0}  =
\left. \frac{d N_{h}^{pp}}{d y} \right|_{y=0}
\left[ (1 - x) N_p/2 + x N_{coll} \right]~,
\end{equation}
where $x=0.11$ for $\sqrt{s}=200$ GeV \cite{PhobosX},
$N_p(b)$ is the number of participants and $N_{coll}(b)$ is the number
of collisions. Both are calculated in the Glauber approach.

Calculating the centrality dependence of the $J/\psi$
suppression, it is convenient to introduce an
{\it effective} participant density in the plane
transverse to the collision axis:
\begin{equation}\label{nstar}
n_p^{*}(\vec{b},\vec{s})=
\left[ (1 - x) n_p(\vec{b},\vec{s}) + 2 x n_c(\vec{b},\vec{s}) \right].
\end{equation}
Here $n_p(\vec{b},\vec{s})$ and $n_c(\vec{b},\vec{s})$ are,
respectively, the densities of nucleon participants
and collisions in the transverse plane:
$N_p(b)=\int d^2s~ n_p(\vec{b},\vec{s})$ and
$N_{coll}(b)=\int d^2s~ n_{coll}(\vec{b},\vec{s})$.
Note that the multiplicity of light-flavored hadrons (\ref{nch_AA})
is proportional to $N_p^{*}(b)=\int d^2s~ n_p^{*}(\vec{b},\vec{s})$.
Motivated by this fact, we assume that the comover {\it density}
in the transverse plane,
which is needed to calculate $S^{co}$, is proportional to $n_p^{*}$.

\begin{figure}[t]
\begin{center}
\includegraphics[width=8.6cm]{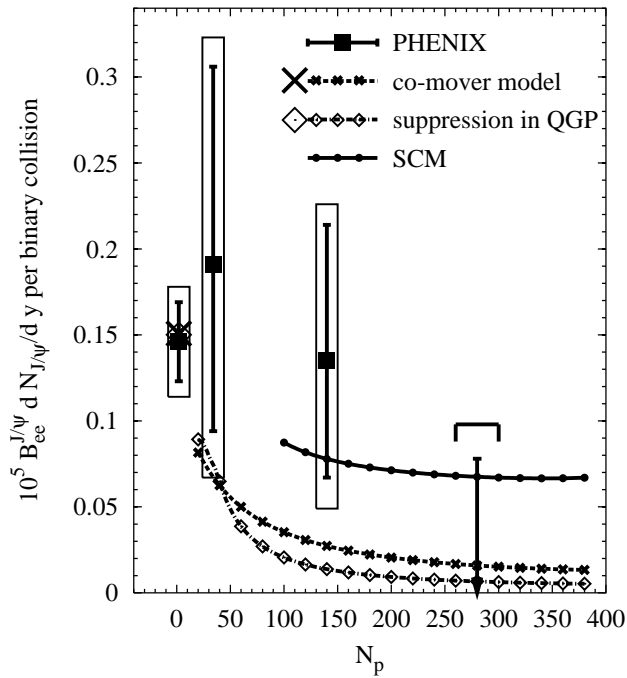}
\caption{The $J/\psi$ multiplicity 
per binary collision in the central unit rapidity interval
at $\sqrt{s}=200$ GeV multiplied by the $J/\psi \rightarrow e^+ e^-$
branching ratio vs the number of nucleon participants.
The most probable values (squares), statistical errors (errorbars),
and quadrature sum of statistical and systematic errors (boxes) 
are shown
for  the $p+p$ and two centralities of Au+Au collisions. 
The 90 \% confidence level (arrow) and its uncertainty (bracket)
are shown for the most central collisions.
\label{Jpsi_RHIC}
 \label{fig1}}
\end{center}
\end{figure}

The cross section of $J/\psi$ production per unit rapidity interval
at midrapidity $\left.  d \sigma^{NN}_{J/\psi}/d y \right|_{y=0}$
is the only free parameter of our fit.
The PHENIX data on the $J/\psi$ multiplicity in
$p+p$ and Au+Au collisions \footnote{The $J/\psi$
multiplicity in $p+p$ is related to the cross section by 
the standard formula
$\left. N_{J/\psi}^{pp}/dy \right|_{y=0}= 1/\sigma_{inel}^{pp}
\left.  d \sigma^{NN}_{J/\psi}/d y \right|_{y=0}$, where $\sigma_{inel}^{pp}$
is the total inelastic $p+p$ cross section.} 
are fitted simultaneously.
The best fit, $\chi^2/ndf = 2.0$ \footnote{Only statistical errors are 
taken into account in the calculation of $\chi^2$.
The most likely 
value of the $J/\psi$ yield in the most central collisions is not reported.
Therefore, this point is not used in our fit procedure.}, is reached
at $B^{J/\psi}_{e^+ e^-} 
\left.  d \sigma^{NN}_{J/\psi}/d y \right|_{y=0}
=4.9 \times
10^{-2}$ $\mu$b ($B^{J/\psi}_{e^- e^+}$ is the branching ratio of
$J/\psi$ decays into electron positron pair).
The result is shown in Fig.~\ref{fig1}.

The comover model has been historically referred to as a 
``hadronic'' model.  One might doubt whether the 
extrapolation of this model
to the RHIC energies is legal.
Indeed, the estimated energy density is extremely high even at 
SPS, so that hadrons can hardly preserve their individuality.
The authors of the comover approach do not insist, however, on its 
hadronic interpretation (see, for instance \cite{nonhadr}).
We do not therefore make any assumptions about the nature of the 
comoving medium. We merely consider the comover model 
as an {\it extreme} scenario, which assumes a gradual increase of the
charmonium suppression with growing energy density, without any abrupt
changes of the absorptive properties of the medium.

The QGP scenario of Ref.\cite{Blaizot} represents the opposite
{\it extreme}: the charmonium suppression sets in, as soon as
the energy density exceeds some threshold value. The excited charmonia,
which contribute about 40\% to the total $J/\psi$ yield, 
are suppressed at lower
energy densities than directly produced $J/\psi$'s.
We have updated the fit \cite{Blaizot}
to the SPS data (new NA50 data \cite{Ramelo} were added)
using corrected values of the parameters,
$\sigma_{abs} = 4.4 \pm 0.5$ mb and
$\sigma^{NN}_{J/\psi}/\sigma^{NN}_{DY}\approx 43.1$,
reported  recently \cite{Cortese}.
Our results are $n_1=2.99$ fm$^{-2}$ and
$n_2=3.86$ fm$^{-2}$. Here
$n_1$ ($n_2$) is the participant density in the
transverse plane, corresponding to the threshold energy density at
which excited charmonia ($J/\psi$'s) are fully suppressed.

Extrapolating to RHIC energies,
one again has to take into account that
the number of produced hadrons per unit rapidity and, consequently,
the energy density of the produced medium grows with the collision energy
and centrality.
Due to the deviation from the wounded
nucleon model (\ref{nch_AA}), the charmonium suppression
sets in, when the {\it effective}
participant density  $n_p^{*}(\vec{b},\vec{s})$ (\ref{nstar}) rather
than the usual $n_p(\vec{b},\vec{s})$ exceeds the threshold values.
The number of secondary hadrons per {\it effective}
participant pair at $\sqrt{s}=200$ is higher
than that at the SPS by a factor of about $1.5$.
The critical energy density at RHIC is reached, therefore,
at lower effective participant density:
$n_{1}^{*} = n_{1}/1.5 \approx
2.0$~fm$^{-2}$ and
$n_{2}^{*}= n_{2}/1.5. \approx
2.6$~fm$^{-2}$.

Similarly as in the comover model, the $J/\psi$
production cross section
is the only free
parameter in the fit of the RHIC data. The minimum 
$\chi^2/ndf = 2.2$ is obtained at
$B^{J/\psi}_{e^+ e^-} \left.  
d \sigma^{NN}_{J/\psi}/d y \right|_{y=0}=5.0 \times 10^{-2}$~$\mu$b.
The best fit of the QGP suppression scenario is also shown in Fig.~1.

As was already noted above, the extrapolation of the standard 
suppression models from SPS to RHIC energies was based on the assumption 
that the value of the normal nuclear absorption cross section $\sigma_{abs}$
does not change essentially with the collision energy. Other viewpoints
are also possible.  
 The $J/\psi$ nuclear suppression mechanism at 
RHIC may be completely different from that at SPS \cite{nns_mod}. 
This does not improve the agreement of the standard suppression models
with the data, if the nuclear suppression becomes stronger at RHIC 
\cite{Capella_nns}. It is not excluded, however, that the nuclear 
suppression may be even weaker \cite{Gerland_nns}. A $J/\psi$ measurements
in $d+Au$ collisions would clarify this point. 

% As it is seen from Fig.~\ref{fig1}, t

% The extrapolation of both, the
% comover and QGP suppression scenarios 
% , to the top RHIC energy
% results in a too strong $J/\psi$ suppression(see Fig.~\ref{fig1}), 
% which is not favored
% by the data. The confidence level of these scenarios is 9\% and
% 3\%, respectively. Initially formed and survived charmonia are
% unlikely to be dominant in A+A collisions at this high energy.
% More likely, charmonium states are formed at a later
% stage of the reaction due to coalescence of $c$'s and $\bar{c}$'s.

Now we will check whether the statistical coalescence
model\cite{BMS,We} can be tolerated by the new data.
In the
SCM \cite{BMS}, the total charm content of the final hadron system
equals the number of $c$ and $\bar{c}$ created at the initial
stage of $A+A$ reaction. Statistical laws control only the
distribution of $c$ and $\bar{c}$ among different hadron states in
terms of the hadron gas (HG) model parameters: temperature $T$,
baryonic chemical potential $\mu_b$ and volume $V$. It appears
that the number of $J/\psi$'s produced by a statistical
coalescence mechanism depends weakly on the thermodynamic
hadronization parameters $T$ and $\mu_{B}$. The $J/\psi$ yield is
mainly defined by the average number of charmed quark-antiquark
pairs $\overline{N}_{c\bar{c}}$ and by the hadronization volume
parameter $V$. If $\overline{N}_{c\bar{c}}$ is not much larger
than unity, a proper account  for the exact charm conservation
becomes essential \cite{We}. This is crucial at  SPS energies,
where $\overline{N}_{c\bar{c}}$ is less than unity, and remains 
essential for moderate centrality in Au+Au collisions at RHIC.

The SCM formula for the {\it total}\ \ ($4 \pi$)\ \ $J/\psi$ yield
that takes into account exact conservation of the number of 
$c\bar{c}$ pairs 
was obtained in Ref.\cite{We}. In real experimental situations, 
however, measurements are performed in a limited rapidity window. 
In the most simple case when
the fraction of charmonia that fall into the relevant rapidity
window does not depend on the centrality, one can merely use
the formula for the total yield multiplied by some factor $\xi < 1$. 
This approach was used \cite{SPS} for studying the SPS data,
where the multiplicity of light hadrons, which determine the
freeze-out volume of the system, are approximately proportional
to the number of nucleon participants $N_p$ at all rapidities.
At RHIC, the situation is different: the {\it total} ($4\pi$)
multiplicity of light hadrons are approximately proportional
to the number of participants, while {\it at midrapidity}, it 
grows faster [see Eq.(\ref{nch_AA})].
The centrality dependence of charmonium production
at different rapidities should, in this case, be also different.
To compare the SCM prediction to the PHENIX data, which are related to
the $J/\psi$ yield at midrapidity $d N_{J/\psi}/dy$, one has to
derive a formula for the $J/\psi$ yield in a {\it finite} rapidity 
interval.

Let $\xi_{\Delta y} < 1$ is the probability that a $c$ quark,
produced in a nucleus-nucleus collision, has rapidity within 
the interval $\Delta y$.
The probability distribution
of the number $k_c$ of $c$ quarks inside the interval $\Delta y$
for events with fixed {\it total} ($4 \pi$) number
$N_{c\bar{c}}$ of $c\bar{c}$ pairs 
is given by the binomial law: 
\begin{equation}
f(k_{c}|N_{c\bar{c}}) =
\frac{N_{c\bar{c}}!}{k_{c}!~(N_{c\bar{c}} - k_{c})!}~\xi_{\Delta
y}^{k_{c}}~ (1-\xi_{\Delta y} )^{N_{c\bar{c}}-k_{c}}.
\end{equation} 
The probability distribution of the number $k_{\bar{c}}$ of
$\bar{c}$'s inside the interval $\Delta y$ is assumed
to be {\it independent}
of $k_c$ \footnote{This differs from 
Ref.\cite{theory1}, where exact equality, $k_{c}=k_{\bar{c}}$,
within the chosen interval $\Delta y$ is assumed. In fact, net charm 
is exactly zero only in the total system. In any finite 
rapidity interval, event-by-event fluctuations with
$k_{c} \not = k_{\bar{c}}$ are possible.}. 
It conforms to the same binomial law.
Event-by-event fluctuations of
the number of $c\bar{c}$ pairs $N_{c\bar{c}}$ created at the
early stage of $A+A$ reaction in independent nucleon-nucleon
collisions, are Poisson distributed:
\begin{equation}
P(N_{c\bar{c}};\overline{N}_{c\bar{c}}) =
\exp \left(- \overline{N}_{c\bar{c}} \right)~
 \frac{\left( \overline{N}_{c\bar{c}} \right)^{N_{c\bar{c}}}}
 {N_{c\bar{c}}~!}.
\end{equation} 
The probability of $c\bar{c}$ coalescence is proportional to the
product of their  
numbers and inversely proportional to the system volume.
The proportionality coefficient depends on the 
thermal densities of the open and hidden charm, and is the same as in 
the case of the total charmonium yield \cite{We}.

The average $J/\psi$
multiplicity at fixed values of $k_{c}$ and $k_{\bar{c}}$ is therefore
given by the formula \cite{Vill}
\begin{equation}\label{fix}
\langle J/\psi \rangle_{k_{c} k_{\bar{c}}}^{\Delta y}  \approx k_{c}
k_{\bar{c}} \frac{n_{J/\psi}^{tot}}{(n_O/2)^2}~ \frac{1}{V_{\Delta
y}}.
\end{equation}
[Deriving Eq.(\ref{fix}) we used the fact that the
thermal number of hadrons with hidden charm  is much smaller than
that with open charm.]
Folding Eq.(\ref{fix}) with the  binomial and Poisson distributions one
gets
\begin{equation}  \label{Jpsi}
\langle J/\psi \rangle^{\Delta y}  \approx \xi_{\Delta y}^2
\overline{N}_{c\bar{c}} \left(\overline{N}_{c\bar{c}} + 1\right)
\frac{n_{J/\psi}^{tot}}{(n_O/2)^2}~ \frac{1}{V_{\Delta y}}~,
\end{equation}
where $n_{O}$ is the thermal density of all open charm hadrons and
$n_{{J/\psi}}^{tot}$ is the total thermal $J/\psi$ density (with 
decay contributions from the excited charmonium states included).
Both $n_{O}$ and $n_{J/\psi}^{tot}$ are calculated in the grand
canonical ensemble with the QGP hadronization parameters
$T,\mu_{B},V_{\Delta y}$ found from fitting the data of
light-flavored \footnote{At RHIC, the strangeness as well as all
other conserved charges, excluding charm, can be safely considered
in the grand canonical ensemble.} hadron yields in the rapidity
interval $\Delta y$. The average number of $c\bar{c}$ pairs
$\overline{N}_{c\bar{c}}$ is,
however, related to their {\it total} ($4 \pi$) yield.

In Au+Au collisions at $\sqrt{s}=200$~GeV, the yield of 
light-flavored hadrons at midrapidity  is fitted within the hadron
gas model with $T=177$~MeV and $\mu_{B}=29$~MeV \cite{BMSR}.
The centrality dependence of the volume is calculated from 
\begin{equation}
V_{\Delta y=1} =
\frac{1}{n_{ch}(T,\mu_{B})}~
1.2~ \frac{d N_{ch}^{\mbox{\footnotesize AuAu}}}{d \eta}~,
% \right|_{\eta=0},
\end{equation}
[the coefficient 1.2 is needed to recalculate the 
number of particles per unit {\it pseudorapidity} ($\eta$) interval 
to that per unit {\it rapidity} ($y$) interval \cite{eta-y}].
Here $n_{ch}$ is the charged hadron density calculated in the HG model.

The average number of the initially produced $c\bar{c}$ 
pairs is proportional to the number of
binary nucleon-nucleon collisions: $\overline{N}_{c\bar{c}}=
N_{coll} (b)~\sigma^{NN}_{c\bar{c}}/\sigma^{NN}_{inel}$.
The charm production cross section, $\sigma^{NN}_{c\bar{c}}$, has been
measured at RHIC by the PHENIX Collaboration \cite{Averbeck}.
% using indirect method based on
% the analysis of single lepton spectra.
The result is consistent with
PYTHIA calculations: $\sigma^{NN}_{c\bar{c}}\approx 650\ \mu$b.

The SCM is applicable only to large systems:
$N_p > 100$ in Pb+Pb at the SPS \cite{BMS,SPS}. Therefore, PHENIX's $p+p$
point and the leftmost Au+Au point, corresponding to $N_p \approx 30$,
cannot be used in the SCM fit procedure. From this reason, we restrict 
ourselves only to a rough estimation of the SCM prediction for the $J/\psi$
yield at midrapidity at the top RHIC energy.

% is not used in our fit procedure. 
% , is, 
% however, retained. The multiplicity
% of secondary hadrons increases with the collision energy. Therefore,
% at RHIC, the system is likely to become sufficiently large
% at smaller number of participants than at SPS.
%It is expected that
%the deviation from the SPS at this point are much smaller than
%the statistical error.

%To make a comparison of the SCM with PHENIX data

We fix the charm production cross section in nucleon-nucleon
collisions at its PYTHIA value,
$\sigma^{NN}_{c\bar{c}} = 650\ \mu$b.
There is no experimental 
data for the value of $\xi_{\Delta y =1}$, but one can roughly 
estimate it assuming approximately the same rapidity 
distribution of the open charm and $J/\psi$'s in $p+p$
collisions. This leads to $\xi_{\Delta y =1} \approx 0.3$.
The charm rapidity distribution in Au+Au collisions
can be broader than in $p+p$ due to rescattering
of $c$ and $\bar{c}$ by sweeping nucleons.
This will not change our result essentially, however. 
The estimation of the total charm production cross section 
is based on the single electron measurement at midrapidity.
Extrapolation to the total phase space has been done 
assuming that the charm rapidity distribution 
does not change from $p+p$ to Au+Au. The charm production rate per 
binary collision at midrapidity was found to be independent 
of the centrality (at least within the present measurement 
accuracy). This means that the value of the total charm 
production cross section 
would grow with the centrality, if there were a broadening of the 
rapidity distribution. Both effects, the decreasing of 
$\xi_{\Delta y =1}$ and the increasing of 
$\sigma^{NN}_{c\bar{c}}$, nearly cancel each other
in Eq. (\ref{Jpsi})
and the prediction of SCM does not change significantly.

The result is shown
in Fig.~\ref{fig1}. The SCM dependence of the $J/\psi$
{\it rapidity density}
per binary collision on the centrality is almost flat
at $N_p \gtrsim 100$, in contrast to the {\it total} $J/\psi$ 
yield, where a $J/\psi$ enhancement is expected \cite{We_RHIC}. 
This difference appears 
because the hadronization volume {\it at midrapidity}, $V_{\Delta y=1}$,
grows with  the centrality faster than the {\it total} volume $V$.

% Note that the value $\xi_{\Delta y =1} \approx 0.18$ is quite 
% reasonable: A smaller 
% value ($\xi_{\Delta y =1} < 0.1$) would mean almost flat open charm 
% distribution within more
% than 10 units of rapidity, while the total rapidity interval
% at $\sqrt{s}=200$ GeV is only about 12 units wide. There is no experimental 
% data for the upper bound of $\xi_{\Delta y =1}$, but one can roughly 
% estimate it assuming 
% that the open charm rapidity distribution in N+N is similar to that of
% $J/\psi$ measured by PHENIX \cite{data}. In Au+Au, the distribution 
% is expected to be approximately the same or even broader due to rescattering
% of $c$ and $\bar{c}$. Therefore $\xi_{\Delta y =1} > 0.3$ is unlikely. 

In conclusion, we have compared predictions of three different models
for $J/\psi$ production at the top RHIC energy $\sqrt{s}=200$ GeV.
None of the models are favored and none is excluded by the data. 
The statistical coalescence model predicts a few times larger 
$J/\psi$ production rate than the standard suppression models.
Hopefully, measurements during the next Au+Au run will be able 
to clarify whether charmonia are formed only at the initial 
stage of the reaction (the standard suppression models) or 
production at the late stage via $c\bar{c}$ coalescence (SCM)
is dominant. A crucial test for the SCM would be a measurement of 
the centrality dependence of $\psi'$ to $J/\psi$ ratio
in Au+Au collisions. It should be 
constant (excluding the peripheral collision region) and equal to 
the value in equilibrium HG, if SCM is valid.

The two standard charmonium suppression models, the gradual 
suppression by comovers and the abrupt suppression by QGP, 
give quite similar predictions. High quality data with small errorbars
are needed to clarify, which of the models describes adequately
the charmonium suppression process, if contribution from 
$c\bar{c}$ coalescence is not significant.

What is the charmonium production 
mechanism at SPS energies? Here both the standard suppression models 
\cite{Capella,comover,Blaizot} 
and SCM \cite{SPS}, as well as their combination \cite{GR} are also able 
to reproduce the data. However, SCM requires an essential (by the 
factor of about $3.5$) enhancement of the open charm in Pb+Pb collisions. 
There is an indirect 
experimental evidence for such an enhancement \cite{NA50open}
and its possible mechanism has been considered \cite{hf_enh}. Still, only 
a direct experimental verification \cite{NA60} can give the final answer.

%\vspace{0.2cm} Conclusions and Discussions. ...

%\begin{acknowledgments}
{\bf Acknowledgments.}
We are grateful to A. Capella for useful suggestions and comments.
Financial support of the Alexander von Humboldt Foundation
and GSI (Germany) is appreciated. 
%\end{acknowledgments}

%\section*{References}
%\begin{harvard}


\begin{thebibliography}{99}

\bibitem{MS}
T.~Matsui and H.~Satz,
%``J / Psi Suppression By Quark - Gluon Plasma Formation,''
Phys.\ Lett.\ B {\bf 178}, 416 (1986).
%%CITATION = PHLTA,B178,416;%%

\bibitem{gago1}
M.~Ga\'zdzicki and M.~I.~Gorenstein,
%``Evidence for statistical production of J/psi mesons in nuclear
% collisions at the CERN SPS,''
Phys.\ Rev.\ Lett.\  {\bf 83}, 4009 (1999).
%[hep-ph/9905515].
%%CITATION = HEP-PH 9905515;%%

\bibitem{BMS}
P.~Braun-Munzinger and J.~Stachel,
%``(Non)thermal aspects of charmonium production and a new look at J/psi
%suppr$
Phys.\ Lett.\ B {\bf 490}, 196 (2000);
%[arXiv:nucl-th/0007059];\\
%%CITATION = NUCL-TH 0007059;%%
%``On charm production near the phase boundary,''
Nucl.\ Phys.\ A {\bf 690}, 119c (2001).
%[arXiv:nucl-th/0012064].
%%CITATION = NUCL-TH 0012064;%%

\bibitem{We}
M.~I.~Gorenstein {\it et al.},
% , A.~P.~Kostyuk, H.~St\"ocker and W.~Greiner,
%``Statistical coalescence model with exact charm conservation,''
Phys.\ Lett.\ B {\bf 509}, 277 (2001);
%[arXiv:hep-ph/0010148];
%%CITATION = HEP-PH 0010148;%%
%``Open charm enhancement in Pb + Pb collisions at SPS,''
J.\ Phys.\ G {\bf 27}, L47 (2001).
%[arXiv:hep-ph/0012015].
%%CITATION = HEP-PH 0012015;%%

\bibitem{SPS}
A.~P.~Kostyuk {\it et al.},
%, M.~I.~Gorenstein, H.~Stocker and W.~Greiner,
%`` Statistical Coalescence Model Analysis Of J / Psi Production In Pb +
%Pb Collisions At 158-A-Gev,''
Phys.\ Lett.\ B {\bf 531},  195 (2002);
%%CITATION = PHLTA,B531,195;%%
%[arXiv:hep-ph/0110269];\\
%%CITATION = HEP-PH 0110269;%%
%``The high E(T) drop of J/psi to Drell-Yan ratio from the statistical c
%anti-c coalescence model,''
J.\ Phys.\ G {\bf 28},  2297 (2002).
%[arXiv:hep-ph/0204180];\\
%%CITATION = HEP-PH 0204180;%%

\bibitem{Bug}
M.~I.~Gorenstein, K.~A.~Bugaev and M.~Gazdzicki,
%``Omega, J/psi and psi' production in nuclear collisions and quark gluon  plasma hadronization,''
Phys.\ Rev.\ Lett.\  {\bf 88}, 132301 (2002).
%%CITATION = HEP-PH 0112197;%%

\bibitem{GR}
L.~Grandchamp and R.~Rapp,
%``Thermal versus direct J/psi production in ultrarelativistic heavy-ion
%collisions
Phys.\ Lett.\ B {\bf 523},  60 (2001);
%[arXiv:hep-ph/0103124];\\
%%CITATION = HEP-PH 0103124;%%
%``Charmonium suppression and regeneration from SPS to RHIC,''
Nucl.\ Phys.\ A {\bf 709}, 415 (2002).
%[arXiv:hep-ph/0205305].
%%CITATION = HEP-PH 0205305;%%

\bibitem{Thews}
R.~L.~Thews, M.~Schroedter and J.~Rafelski,
%``Enhanced J/psi production in deconfined quark matter,''
Phys.\ Rev.\ C {\bf 63}, 054905 (2001).

\bibitem{data_prel}
A.~D.~Frawley  [PHENIX Collaboration],
%``J/psi $\to$ e e and J/psi $\to$ mu mu measurements in Au Au 
% and p p collisions at s(NN)**(1/2) = 200-GeV,''
Nucl.\ Phys.\ A {\bf 715}, 687 (2003).
%%CITATION = NUCL-EX 0210013;%%

\bibitem{theory1}
A.~Andronic {\it et al.},
%``Statistical hadronization of charm at SPS, RHIC and LHC,''
Nucl.\ Phys.\ A {\bf 715}, 529 (2003);
%%CITATION = NUCL-TH 0209035;%%
Phys.\ Lett.\ B {\bf 571}, 36 (2003).
%%CITATION = NUCL-TH 0303036;%%

\bibitem{theory2}
L.~Grandchamp and R.~Rapp,
%``Two-component approach to J/psi production in high-energy heavy-ion 
%  collisions,''
Nucl.\ Phys.\ A {\bf 715}, 545 (2003).
%%CITATION = HEP-PH 0209141;%%

\bibitem{theory3}
E.~L.~Bratkovskaya, W.~Cassing and H.~Stocker,
%``Open charm and charmonium production at RHIC,''
Phys.\ Rev.\ C {\bf 67}, 054905 (2003).
%%CITATION = NUCL-TH 0301083;%%

\bibitem{data}
S.~S.~Adler {\it et al.}  [PHENIX Collaboration],
%``J/psi production in Au - Au collisions at s**(NN)(1/2) = 200-GeV at the  
% Relativistic Heavy Ion Collider,'' 
Phys.\ Rev.\ C (to be published),
arXiv:nucl-ex/0305030.
%%CITATION = NUCL-EX 0305030;%%

\bibitem{Kharzeev}
D.~Kharzeev {\it et al.},
%, C.~Lourenco, M.~Nardi and H.~Satz,
%``A quantitative analysis of charmonium suppression in nuclear  collisions,''
Z.\ Phys.\ C {\bf 74}, 307 (1997).
%[hep-ph/9612217].
%%CITATION = HEP-PH 9612217;%%

\bibitem{Capella}
N.~Armesto and A.~Capella,
%``A quantitative reanalysis of J/psi suppression in nuclear collisions,''
Phys.\ Lett.\ B {\bf 430}, 23 (1998);\\
%[arXiv:hep-ph/9705275];\\
%%CITATION = HEP-PH 9705275;%%
N.~Armesto, A.~Capella and E.~G.~Ferreiro,
%``Charmonium suppression in lead lead collisions: Is there a break in the
% J/psi cross-section?,''
Phys.\ Rev.\ C {\bf 59}, 395 (1999).
%[hep-ph/9807258];\\
%%CITATION = HEP-PH 9807258;%%

\bibitem{Cortese}
P.~Cortese {\it et al.}  [NA50 Collaboration],
%``Charmonia Absorption In P A Collisions At The Cern Sps: 
% Results And Implications For Pb Pb Interactions,''
Nucl.\ Phys.\ A {\bf 715}, 679 (2003).
%%CITATION = NUPHA,A715,679;%%

\bibitem{Gerland}
L.~Gerland {\it et al.},
%, L.~Frankfurt, M.~Strikman, H.~Stocker and W.~Greiner,
%``J/psi production, chi polarization and color fluctuations,''
Phys.\ Rev.\ Lett.\  {\bf 81},  762 (1998).
%%CITATION = NUCL-TH 9803034;%%

\bibitem{comover}
C.~Spieles {\it et al.},
%R.~Vogt, L.~Gerland, S.~A.~Bass, M.~Bleicher,
%H.~St\"ocker and W.~Greiner,
%``Modelling J/psi production and absorption in a microscopic
% nonequilibrium approach,''
Phys.\ Rev.\ C {\bf 60},  054901 (1999);\\
%[hep-ph/9902337];\\
%%CITATION = HEP-PH 9902337;%%
J.~Geiss {\it et al.},
%C.~Greiner, E.~L.~Bratkovskaya, W.~Cassing and U.~Mosel,
%``Charmonium suppression with c anti-c dissociation by strings,''
Phys.\ Lett.\ B {\bf 447}, 31 (1999);\\
%[nucl-th/9803008];\\
%%CITATION = NUCL-TH 9803008;%%
D.~E.~Kahana and S.~H.~Kahana,
%``J/psi suppression in heavy ion collisions at the CERN SPS,''
Prog.\ Part.\ Nucl.\ Phys.\ {\bf 42}, 269 (1999).
%%CITATION = PPNPD,42,269;%%

\bibitem{Blaizot}
J.~P.~Blaizot and J.~Y.~Ollitrault,
%``J/psi suppression in Pb Pb collisions: A hint of quark-gluon plasma
% production?,''
Phys.\ Rev.\ Lett.\  {\bf 77}, 1703 (1996);\\
%[arXiv:hep-ph/9606289].
%%CITATION = HEP-PH 9606289;%%
J.~P.~Blaizot, P.M.~Dinh and J.~Y.~Ollitrault,
%``Transverse energy fluctuations and the pattern of J/psi suppression in
% Pb Pb collisions,''
{\it ibid.}  {\bf 85}, 4012 (2000).
%[arXiv:nucl-th/0007020].
%%CITATION = NUCL-TH 0007020;%%

\bibitem{Capella_new}
A.~Capella and D.~Sousa,
%``New J/psi suppression data and the comovers interaction model,''
nucl-th/0303055.
%%CITATION = NUCL-TH 0303055;%%

\bibitem{Ramelo}
L.~Ramello {\it et al.}  [NA50 Collaboration],
%``Results On Leptonic Probes From Na50,''
Nucl.\ Phys.\ A {\bf 715}, 243 (2003).
%%CITATION = NUPHA,A715,243;%%

\bibitem{KhN}
D.~Kharzeev and M.~Nardi,
%``Hadron production in nuclear collisions at RHIC and high density QCD,''
Phys.\ Lett.\ B {\bf 507}, 121 (2001).
%[arXiv:nucl-th/0012025].

\bibitem{PhobosX}
B.~B.~Back {\it et al.}  [PHOBOS Collaboration],
%``Centrality dependence of the charged particle multiplicity near
% mid-rapidity in Au + Au collisions at s(NN)**(1/2) = 130-GeV and  200-GeV,''
Phys.\ Rev.\ C {\bf 65}, 061901 (2002).
%[arXiv:nucl-ex/0201005].
%%CITATION = NUCL-EX 0201005;%%

\bibitem{nonhadr}
A.~Capella,
%``Microscopic models of heavy ion interactions,''
arXiv:hep-ph/0305196.
%%CITATION = HEP-PH 0305196;%%

\bibitem{nns_mod}
L.~Gerland, L.~Frankfurt, M.~Strikman, H.~Stocker and W.~Greiner,
%``Suppression of quarkonium production in heavy ion collisions at RHIC  
% and LHC,''
J.\ Phys.\ G {\bf 27}, 695 (2001);\\
%%CITATION = NUCL-TH 0009008;%%
B.~Kopeliovich, A.~Tarasov and J.~H\"ufner,
%``Coherence phenomena in charmonium production off nuclei at the energies  
% of RHIC and LHC,''
Nucl.\ Phys.\ A {\bf 696}, 669 (2001);\\
%%CITATION = HEP-PH 0104256;%%
A.~Capella,
%``Coherence Effects in Charmonium Production off Nuclei : Consequences for 
% J/\psi Suppression,''
arXiv:nucl-th/0207049.
%%CITATION = NUCL-TH 0207049;%%

\bibitem{Capella_nns}
A.~Capella and D.~Sousa,
%``Charged multiplicities and J/psi suppression at SPS and RHIC energies,''
arXiv:nucl-th/0110083.
%%CITATION = NUCL-TH 0110083;%%

\bibitem{Gerland_nns}
L.~Gerland, L.~Frankfurt, M.~Strikman, H.~Stocker and W.~Greiner,
%``Suppression of quarkonium production in heavy ion collisions at RHIC  
% and LHC,''
J.\ Phys.\ G {\bf 27}, 695 (2001).
%%CITATION = NUCL-TH 0009008;%%

\bibitem{Vill}
A.~P.~Kostyuk,
%``Statistical hadronization of charm in heavy ion collisions,''
arXiv:hep-ph/0306123.
%%CITATION = HEP-PH 0306123;%%

\bibitem{eta-y}
K.~Adcox {\it et al.}  [PHENIX Collaboration],
%``Centrality dependence of charged particle multiplicity in Au Au
% collisions at s(N N)**(1/2) = 130-GeV,''
Phys.\ Rev.\ Lett.\  {\bf 86}, 3500 (2001).
%[arXiv:nucl-ex/0012008].
%%CITATION = NUCL-EX 0012008;%%

\bibitem{BMSR}
P.~Braun-Munzinger, K.~Redlich and J.~Stachel,
%``Particle production in heavy ion collisions,''
nucl-th/0304013.
%%CITATION = NUCL-TH 0304013;%%

\bibitem{Averbeck}
R.~Averbeck  [PHENIX Collaboration],
%``Single leptons from heavy-flavor decays at RHIC,''
Nucl.\ Phys.\ A {\bf 715}, 695 (2003).
%%CITATION = NUCL-EX 0209016;%%

\bibitem{We_RHIC}
M.~I.~Gorenstein {\it et al.}
%, A.~P.~Kostyuk, L.~McLerran, H.~Stocker and W.~Greiner,
%``Open And Hidden Charm Production In Heavy-Ion Collisions At
% Ultrarelativistic Energies,''
J.\ Phys.\ G {\bf 28}, 2151 (2002).
%%CITATION = JPHGB,G28,2151;%%
%[arXiv:hep-ph/0012292].
%%CITATION = HEP-PH 0012292;%%

\bibitem{NA50open}
M.~C.~Abreu {\it et al.}  [NA38 and NA50 Collaborations],
%``Dimuon and charm production in nucleus nucleus collisions at the CERN-SPS,''
Eur.\ Phys.\ J.\ C {\bf 14}, 443 (2000).
%%CITATION = EPHJA,C14,443;%%

\bibitem{hf_enh}
A.~P.~Kostyuk, M.~I.~Gorenstein and W.~Greiner,
%``Heavy flavor enhancement as a signal of color deconfinement,''
Phys.\ Lett.\ B {\bf 519}, 207 (2001).
%[hep-ph/0103057].
%%CITATION = HEP-PH 0103057;%%

\bibitem{NA60}
B.~Lenkeit {\it et al.} [NA60-Collaboration],
%``Prompt dimuons and D meson production in heavy-ion collisions at the  SPS,''
arXiv:nucl-ex/0108015.
%%CITATION = NUCL-EX 0108015;%%


\end{thebibliography}
\end{document}